\documentstyle[12pt]{article}

\newcommand{\mib}[1]{\mbox{\boldmath $#1$}}

\begin{document}
%\begin{center}
%Bloch theorem and Spontaneous Current
%\end{center}
\begin{center}
{\Large\bf On the Bloch Theorem}
\medskip

{\Large\bf Concerning Spontaneous Electric 
Current\footnote{Submitted to J. Phys. Soc. Jpn.}}
\vskip1cm
{\large Yoji {\sc Ohashi} and Tsutomu {\sc Momoi}}\\
\bigskip

{\it Institute of Physics, University of Tsukuba, Tsukuba, Ibaraki 305, 
Japan.}
\bigskip

(May 9, 1996)
\end{center}

\begin{abstract}
\normalsize
\noindent
We study the Bloch theorem which states absence of the spontaneous current 
in interacting electron systems. This theorem is shown to be still applicable 
to the system with the magnetic field induced by the electric current. 
Application to the spontaneous surface current is also examined in detail. 
Our result excludes the possibility of the recently proposed $d$-wave 
superconductivity having the surface flow and finite total current. 
\end{abstract} 
\section{Introduction}
\hspace*{\parindent}
Although the symmetry of the order parameter in high-$T_{\rm c}$ 
superconductivity has not been identified yet, we have some experimental 
evidences which indicate the $d$-wave superconducting 
state.\cite{rf1,rf2,rf3,rf4,rf5,rf6,rf7,rf8} Since this type of 
the symmetry is qualitatively different from the conventional $s$-wave 
superconductivity, we can expect some new phenomena in high-$T_{\rm c}$ 
superconductivity.
\par
Recently, some researchers pointed out the possibility of the $d$-wave 
superconductivity with broken time-reversal symmetry, which results from 
the $s+{\rm i}d$-wave state near the surface.\cite{rf9,rf10,rf11} 
In addition, it was discussed that this state has a finite electric 
surface current.\cite{rf12,rf13} In this case, we should remark that 
this state has the following properties:
\begin{itemize}
\item[(1)] The current flows without any external field, i.e., 
it appears {\it spontaneously}.
\item[(2)] The total current is also finite.\cite{rf12}
\end{itemize}
On the other hand, the Bloch theorem states {\it absence of the 
stationary current in the ground state with no external 
field}.\cite{rf14,rf15,rf16,rf17,rf18} Then the above surface current 
seems inconsistent with this theorem. 
\par
Motivated by this discrepancy, we study the Bloch theorem and its 
application to this state. The Bloch theorem has been examined so far 
within the framework of the first quantization, which seems disadvantageous 
to superconductivity. Furthermore, the theorem in case with the magnetic 
field induced by the spontaneous current has not been examined in detail. 
For these reasons, we present the second quantized version of the Bloch 
theorem, which can directly cover the superconductivity, and also show 
its extension to the system with the induced magnetic field. 
\par
This paper is organized as follows: In ${\S}2$, we examine the Bloch 
theorem neglecting the magnetic field induced by the current, because 
the spontaneous surface current is reported in this 
situation.\cite{rf12,rf13} Effects of the induced magnetic field are 
discussed in ${\S}3$, which is followed by the summary in ${\S}4$.

\section{Bloch Theorem}
\hspace*{\parindent}
In this section, we first of all rederive the Bloch theorem from general 
viewpoint. Possibility of the spontaneous surface current is then discussed 
on the basis of this theorem. 

\subsection{Definitions}
\hspace*{\parindent}
 We start with the case of the chargeless (neutral) fermion, in which 
the current does not induce magnetic field. We also assume the absence 
of the external magnetic or electric field. 
Then, the Hamiltonian is given by
\begin{eqnarray}
H&=&\sum_\sigma\int{\rm d}{\mib r}
\Psi_\sigma^\dagger({\mib r})
[({{\hat {\mib p}}^2 \over 2m}-\mu)+V({\mib r})]
\Psi_\sigma({\mib r})\nonumber\\
&+&{1 \over 2}
\sum_{\sigma,\sigma'}
\int{\rm d}{\mib r}{\rm d}{\mib r}'
\Psi_\sigma^\dagger({\mib r})
\Psi_{\sigma'}^\dagger({\mib r}')
U({\mib r}-{\mib r}')
\Psi_{\sigma'}({\mib r}')
\Psi_\sigma({\mib r}).
\label{eq.2.1}
\end{eqnarray} 
Here ${\hat {\mib p}}\equiv\bigtriangledown/{\rm i}$, while $m$ and 
$\mu$ represent the mass of an electron and the chemical potential, 
respectively. $\Psi_\sigma({\mib r})$ means the electron field operator 
with spin-$\sigma$, and $U({\mib r}-{\mib r}')$ is the electron-electron 
interaction, in which the pairing interaction can be included. 
The lattice and impurity potentials are represented by $V({\mib r})$. 
(To describe the presence of the surface, we put the potential $V({\mib r})$ 
infinity outside the sample.)
\par
We also introduce the current density operator
\begin{equation}
{\hat {\mib J}}({\mib r})={1 \over 2m}\sum_{\sigma}
[\Psi_\sigma^\dagger({\mib r}){\hat {\mib p}}\Psi_\sigma({\mib r})
-
({\hat {\mib p}}\Psi_\sigma^\dagger({\mib r}))\Psi_\sigma({\mib r})].
\label{eq.2.2}
\end{equation}

At this stage, we comment on the reason why this paper uses the second 
quantization in contrast to the previous papers,\cite{rf14,rf17} which 
prove the Bloch theorem in the language of the first quantization. 
When superconductivity is examined, appropriate infinitesimal symmetry 
breaking field is formally necessary in order to select a special state 
from various superconducting states.
In this case, the corresponding fictitious field Hamiltonian ($H_{\rm FF}$) 
has a number non-conserving form, and hence it is difficult to write down 
within the first quantization. We thus use the second quantization
 in order to avoid this problem. Indeed, $H_{\rm FF}$ 
for the superconductivity is given by
\begin{equation}
H_{\rm FF}=
\int{\rm d}{\mib r}{\rm d}{\mib r}'
[
h({\mib r},{\mib r}')\Psi_\uparrow^\dagger({\mib r})
\Psi_\downarrow^\dagger({\mib r}')
+h.c.],
\label{eq.2.4}
\end{equation}
where $h({\mib r},{\mib r}')$ is the infinitesimal symmetry breaking field. 
In what follows, though $H_{\rm FF}$ is not written explicitly 
in eq.~(\ref{eq.2.1}), the symmetry breaking field is always put zero 
after the system size is set infinity.

\subsection{Ground state}
\hspace*{\parindent}
Firstly, we examine the system at $T=0$. We assume that the ground state 
$|\psi\rangle$ has a finite total current ${\mib J}_\psi$,
\begin{equation}
{\mib J}_{\psi}
=\int{\rm d}{\mib r}{\mib J}_\psi({\mib r})
=\int{\rm d}{\mib r}\langle\psi|{\hat {\mib J}}({\mib r})|\psi\rangle
=\langle\psi|\sum_\sigma\int{\rm d}{\mib r}
\Psi_\sigma^\dagger({\mib r})
{{\hat {\mib p}} \over m}
\Psi_\sigma({\mib r})
|\psi\rangle
\ne 0,
\label{eq.2.3}
\end{equation}
where ${\mib J}_\psi({\mib r})$ denotes a local current. 
In the following discussion, we show that this assumption of 
finite ${\mib J}_\psi$ leads to a contradiction.
\par
Let us consider a trial state\cite{rf14}
\begin{equation}
|\phi\rangle\equiv
{\rm exp}\{{\rm i}\delta{\mib P}\cdot\sum_\sigma\int{\rm d}{\mib r}
\Psi_\sigma^\dagger({\mib r})
{\mib r}
\Psi_\sigma({\mib r})
\}
|\psi\rangle,
\label{eq.2.5}
\end{equation}
and compare $\langle\phi|H|\phi\rangle$ with the ground state energy, 
$\langle\psi|H|\psi\rangle$. Physically, $\delta{\mib P}$ 
in eq.~(\ref{eq.2.5}) is regarded as a fluctuation applied to $|\psi\rangle$. 
If $|\psi\rangle$ is the true ground state, it must be stable for 
arbitrary $\delta{\mib P}$, i.e., 
$\langle\psi|H|\psi\rangle\le\langle\phi|H|\phi\rangle$. 
\par
It is easily found that each of the (chemical) potential 
and the interaction terms gives the same expectation value in both states. 
On the other hand, the kinetic term gives
\begin{equation}
\langle\phi|\sum_\sigma\int{\rm d}{\mib r}\Psi_\sigma^\dagger({\mib r})
{{\hat {\mib p}}^2 \over 2m}\Psi_\sigma({\mib r})|\phi\rangle
=
\langle\psi|\sum_\sigma\int{\rm d}{\mib r}\Psi_\sigma^\dagger({\mib r})
{({\hat {\mib p}}+\delta{\mib P})^2 \over 2m}\Psi_\sigma({\mib r})|\psi\rangle.
\label{eq.2.6}
\end{equation}
Using eq.~(\ref{eq.2.3}), we obtain
\begin{equation}
\langle\phi|H|\phi\rangle=\langle\psi|H|\psi\rangle
+\delta{\mib P}\cdot{\mib J}_\psi+{N \over 2m}\delta P^2,
\label{eq.2.7}
\end{equation}
where $N$ represents the number of electrons, 
\begin{equation}
N=\langle\psi|\sum_\sigma\int{\rm d}{\mib r}\Psi_\sigma^\dagger({\mib r})
\Psi_\sigma({\mib r})|\psi\rangle.
\label{eq.2.8}
\end{equation}
Now take $\delta{\mib P}$ opposite to ${\mib J}_\psi$ in direction 
and small enough so that the third term in the right hand side 
in eq.~(\ref{eq.2.7}) is negligible compared with the second one. 
Then we find that $|\phi\rangle$ has a lower energy than $|\psi\rangle$,
\begin{equation}
\langle\phi|H|\phi\rangle=\langle\psi|H|\psi\rangle
-|\delta{\mib P}\cdot{\mib J}_\psi|<\langle\psi|H|\psi\rangle.
\label{eq.2.9}
\end{equation}
Though $|\phi\rangle$ is not always the eigenstate of $H$, 
the variational principle guarantees that $|\psi\rangle$ cannot be 
the ground state. This shows that {\it the ground state of the 
chargeless fermions cannot have finite total current.} 
We thus obtain the second quantized version of the Bloch theorem, 
which can cover the superconducting system. Moreover, we find that 
the theorem holds irrespective to the detail of the potential. 
We will use this result when the surface current is studied in section 2.4.
\par
We can extend this theorem to the system with magnetic impurities as well. 
Consider
\begin{equation}
H_{\rm mag}=\sum_{\sigma,\sigma'}
\int{\rm d}{\mib r}
\sum_iu_i({\mib r}-{\mib R_i})
{\mib S}_i\cdot
\Psi_\sigma^\dagger({\mib r})
{\mib \sigma}_{\sigma,\sigma'}
\Psi_{\sigma'}({\mib r}),
\label{eq.2.10}
\end{equation}
where $u_i$ is the potential of an impurity at ${\mib R}_i$ with 
spin ${\mib S}_i$, and ${\mib \sigma}$ represents the Pauli matrix. 
Then because of
\begin{equation}
[H_{\rm mag},~~
{\rm exp}\{{\rm i}\delta{\mib P}\cdot\sum_\sigma\int{\rm d}{\mib r}
\Psi_\sigma^\dagger({\mib r})
{\mib r}
\Psi_\sigma({\mib r})
\}
]=0,
\label{eq.2.11}
\end{equation}
we obtain $\langle\phi|H_{\rm mag}|\phi\rangle
=\langle\psi|H_{\rm mag}|\psi\rangle$; the Bloch theorem is still valid 
even if magnetic impurities are present.

\subsection{Finite temperature}
\hspace*{\parindent}
In the next step, we examine the thermodynamic stability of 
the spontaneous current. Introduce an orthonormal complete set of 
eigenstates $\{|\psi_i\rangle\}$ and define the density matrix 
${\hat \rho}_\psi$ which gives the lowest free energy at $T$:
\begin{equation}
{\hat \rho}_\psi\equiv
\sum_i|\psi_i\rangle w_i \langle\psi_i|.
\label{eq.2.14}
\end{equation}
Here $w_i$ is the statistical weight which satisfies 
$\sum_iw_i=1$ and $0\le w_i\le 1$. (In many cases, the weight $w_i$ is set as 
$e^{-E_i/T}/{\rm Tr}[e^{-H/T}]$, where $E_i$ represents the energy 
of the eigenstate $|\psi_i\rangle$.) 
We again assume that ${\hat \rho}_\psi$ gives a finite total current, 
${\mib J}_\psi$:
\begin{equation}
{\mib J}_{\psi}
=\int{\rm d}{\mib r}{\mib J}_\psi({\mib r})
=\int{\rm d}{\mib r}{\rm Tr}[{\hat \rho}_\psi{\hat {\mib J}}({\mib r})]
={\rm Tr}[{\hat \rho}_\psi\sum_\sigma\int{\rm d}{\mib r}
\Psi_\sigma^\dagger({\mib r})
{{\hat {\mib p}} \over m}
\Psi_\sigma({\mib r})]
\ne 0.
\label{eq.2.15}
\end{equation}
In the following, we show that ${\hat \rho}_\psi$ is always 
accompanied by another density matrix which gives a lower free energy, 
i.e., ${\hat \rho}_\psi$ cannot describe the most stable state.
\par
We consider a trial density matrix created by the operation of 
the exponential operator in eq.~(\ref{eq.2.5}):
\begin{equation}
{\hat \rho}_\phi\equiv
\sum_i|\phi_i\rangle w_i \langle\phi_i|,
\label{eq.2.17}
\end{equation}
where
\begin{equation}
|\phi_i\rangle\equiv
{\rm exp}\{{\rm i}\delta{\mib P}\cdot\sum_\sigma\int{\rm d}{\mib r}
\Psi_\sigma^\dagger({\mib r})
{\mib r}
\Psi_\sigma({\mib r})
\}
|\psi_i\rangle.
\label{eq.2.16}
\end{equation}
It is found that $\{|\phi_i\rangle\}$ also forms an orthonormal complete set. 
\par
Since ${\hat \rho}_\psi$ and ${\hat \rho}_\phi$ have the same 
statistical weight, the entropy is also the same for these states,
\begin{equation}
S_\phi\equiv
-{\rm Tr}[{\hat \rho}_\phi\log{\hat \rho}_\phi]=
-\sum_iw_i\log w_i
=
-{\rm Tr}[{\hat \rho}_\psi\log{\hat \rho}_\psi]\equiv
S_\psi.
\label{eq.2.18}
\end{equation}
On the other hand, the expectation values of the energy of ${\hat \rho}_\phi$ 
($E_\phi={\rm Tr}[{\hat \rho}_\phi H]$) and ${\hat \rho}_\psi$ 
($E_\psi={\rm Tr}[{\hat \rho}_\psi H]$) are relating to each other in the form
\begin{eqnarray}
E_\phi&=&
\sum_iw_i\langle\phi_i|H|\phi_i\rangle
=
\sum_iw_i\langle\psi_i|H|\psi_i\rangle
+\sum_iw_i
\langle\psi_i|
\sum_\sigma\int{\rm d}{\mib r}\Psi_\sigma^\dagger({\mib r})
(
{\delta{\mib P}\cdot{\hat {\mib p}} \over m}
+
{\delta P^2 \over 2m}
)
\Psi_\sigma({\mib r})|\psi_i\rangle
\nonumber\\
&=&
E_\psi+\delta{\mib P}\cdot{\mib J}_\psi+{N \over 2m}\delta P^2.
\label{eq.2.19}
\end{eqnarray}
{}From eqs.~(\ref{eq.2.18}) and (\ref{eq.2.19}), we can compare 
the free energy for ${\hat \rho}_\phi$ $(\equiv F_\phi)$ with that 
for ${\hat \rho}_\psi$ $(\equiv F_\psi)$:
\begin{equation}
F_\phi=E_\phi-TS_\phi=
F_\psi+\delta{\mib P}\cdot{\mib J}_\psi+{N \over 2m}\delta P^2.
\label{eq.2.20}
\end{equation}
For a $\delta{\mib P}$ being small so that the term with $O(\delta P^2)$ 
in eq.~(\ref{eq.2.20}) can be neglected and having an opposite direction 
to ${\mib J}_\psi$, the free energy of the trial state, $F_\phi$, 
is smaller than $F_\psi$ for arbitrary statistical weights. 
(Even if we put ${\hat \rho}_\psi$ the Gibbs state which must give 
the lowest free energy, $F_\psi>F_\phi$ holds.) In conclusion, 
{\it the state with a finite total current is not thermodynamically stable}.
\par
Before ending this section, we briefly note Bohm's study for 
a finite current at $T\ne 0$.\cite{rf14} In his proof, 
the off-diagonal matrix elements of $H$ for $\{|\phi_i\rangle\}$ are 
dropped by using the properties that (i) they are $O(\delta{\mib P})$ and 
(ii) from the perturbative analysis, they are found to affect the energy 
at most $O(\delta P^2)$. However, if there is degeneracy in the diagonal 
elements and, for example, 
$\langle\phi_i|H|\phi_i\rangle=\langle\phi_j|H|\phi_j\rangle~(i\ne j)$ 
is realized, (i,j)-element may affect the energy in $O(\delta{\mib P})$. 
Hence this off-diagonal element cannot be neglected within the perturbation 
theory. On the other hand, since our proof does not rely on the perturbation 
theory, we do not meet this problem.

\subsection{Application to the spontaneous surface current}
\hspace*{\parindent}
Recently, the spontaneous surface current has been discussed 
in unconventional superconductivity. We examine this problem on the basis 
of the Bloch theorem. 
\par
Let us consider a semi-infinite system as shown in Fig.~1: 
Presence of the surface is described by putting $V(x\le 0)=\infty$. 
The surface may have a roughness. 
At the end edges of the system, we take the open boundary condition, 
so that we can choose arbitrary small $\delta{\mib P}$. 
(The system size is finally put infinity.)
\par
Suppose the presence of the spontaneous surface current as shown in Fig.~2(a). 
Then, what does the Bloch theorem state about this situation? 
The answer is that {\it there must be another counter current 
inside the system which cancels the surface one as a whole} (Fig.~2(b)). 
Indeed, we can find an example of this situation in the domain wall problem 
in unconventional superconductivity.\cite{rfb,rfc} In this example, 
the local surface currents exist near both the sides of the wall. 
They, however, flow toward opposite directions to each other, 
thereby satisfying the Bloch theorem. 
At this stage, we emphasize that the presence of the counter flow 
purely results from the Bloch theorem and is irrespective to the origin 
of the surface current. \par
In the next step, we discuss some specific systems 
in which the spontaneous surface currents are proposed.
\begin{itemize}
\item[(1)]
As noted in the introduction, the surface current which is also finite 
in total was proposed in a $d$-wave superconductivity with broken 
time-reversal symmetry near the surface.\cite{rf12,rf13} 
Since the detail of the state, e.g., whether the time-reversal symmetry 
is broken or not, does not affect the Bloch theorem, 
we immediately find that it is not stable at all temperatures.  
\item[(2)]
Sigrist {\it et. al.} also reported the surface current in an unconventional 
superconductivity with broken time-reversal symmetry.\cite{rfb,rfc} 
They showed that the surface flow is canceled by the screening current 
inside the system resulting from the Meissner effect. 
Apparently, their results do not contradict the Bloch theorem. 
(Exactly speaking, the Meissner effect needs the charge of electron, 
which is neglected in this section. 
In $\S 3$, the Bloch theorem is shown to be still valid in the presence of 
the charge.) However, we have to note that the Meissner effect 
is not intrinsic origin of the back flow. 
The counter current has to appear even for the chargeless electrons 
where the Meissner effect does not work; the role of the Meissner effect 
is merely determining the distribution of the local current so as 
to exclude the magnetic field from the superconductor. 
\end{itemize}

Before going to the next section, we comment on the boundary condition 
at the end edges of the system. 
One may think that there is not stationary current under the open boundary 
condition (isolated system); however, when the thermodynamic limit is taken 
with appropriate symmetry breaking field, we can safely reach the bulk 
system in which the stationary current can flow. 
Furthermore, it is expected that the detail of the boundary condition 
does not affect the presence of the surface current. 
Consequently, although we cannot examine the cases with arbitrary boundary 
conditions, the statement that the surface current state is not stable 
is believed to describe the property of the real system.

\subsection{Possibility of the spontaneous circulating current}
\hspace*{\parindent}
We examine the possibility of the circulating flow. 
In a real system, a circuit as shown in Fig.~3(a) is necessary in order 
to maintain the current. 
In this case, the total current passing through the line "A" in Fig.~3(a) 
is zero, which looks escaping from the Bloch theorem. 
Then, is the spontaneous current possible when the circuit is taken into 
account?
\par
Bohm proved for a large ring that the ground state must be in zero total 
angular momentum.\cite{rf14} Recently, Vignale extended Bohm's proof 
to systems with toroidal geometry.\cite{end2} The outline of Bohm's proof 
is as follows: Consider a large but thin superconducting ring with 
the width $L\ll R$ (Fig.~3(b)). 
The ground state $|\psi\rangle$ is assumed to have a finite total angular 
momentum as a result of the circulating current. The trial state is
\begin{equation}
|\phi\rangle\equiv
{\rm exp}\{{\rm i}n\sum_\sigma\int{\rm d}{\mib r}
\Psi_\sigma^\dagger({\mib r})
\theta
\Psi_\sigma({\mib r})
\}
|\psi\rangle.
\label{eq.A.1}
\end{equation}
Because of the single-value condition of $|\phi\rangle$, $n$ must be 
an integer. We find
\begin{eqnarray}
\langle\phi|H|\phi\rangle=\langle\psi|H|\psi\rangle&+&
\langle\psi|\sum_\sigma\int{\rm d}{\mib r}
\Psi_\sigma^\dagger({\mib r}){n{\hat L_z} \over mr^2}
\Psi_\sigma({\mib r})|\psi\rangle
\nonumber\\
\cr
&+&
\langle\psi|\sum_\sigma\int{\rm d}{\mib r}
\Psi_\sigma^\dagger({\mib r}){n^2\over 2mr^2}
\Psi_\sigma({\mib r})
|\psi\rangle,
\label{eq.A.2}
\end{eqnarray}
where ${\hat L}_z$ is the $z$-component of the angular momentum operator. 
When the total current is $J_\psi$, the angular momentum is estimated as 
$\langle L_z\rangle\simeq mRJ_\psi$. 
Then the second and third terms in eq.~(\ref{eq.A.2}) are evaluated as 
$nJ_\psi/R$ and $n^2N/(2mR^2)$, respectively. 
Now, let us examine the case with $J_\psi=O(R)$. 
Because of $N=O(R)$, the third term in eq.~(\ref{eq.A.2}) becomes negligible 
compared with the second one for large $R$. 
Thus, when we choose $n$ so as to be $nJ_\psi<0$, it is found that 
$|\phi\rangle$ has a lower energy than the assumed ground state. 
Consequently, the state that has a finite angular momentum due 
to a circulating current with $O(R)$ cannot be the ground state.
\par
In conclusion, {\it even if the circuit is taken into account,
the spontaneous current which is proportional to the size along the circuit 
is not realized in the bulk system.}

\section{Extension of the Bloch Theorem to the Case with 
Induced Magnetic Field}
\hspace*{\parindent}
In this section, we examine effects of the charge. 
Since the real electron has the charge $e$, the electric current 
induces magnetic field. 
Then, one may expect that the spontaneous current can be stabilized 
by the induced magnetic field. 
This section is devoted to study this possibility.

\subsection{Definitions}
\hspace*{\parindent}
We summarize the definitions and equations used in this section. 
The Hamiltonian and the current density operator are given by
\begin{eqnarray}
H&=&\sum_\sigma\int{\rm d}{\mib r}
\Psi_\sigma^\dagger({\mib r})
\bigl[[{({\hat {\mib p}}-e{\mib A}({\mib r}))^2 \over 2m}-\mu]
+V({\mib r})\bigr]\Psi_\sigma({\mib r})\nonumber\\
&+&{1 \over 2}
\sum_{\sigma,\sigma'}
\int{\rm d}{\mib r}{\rm d}{\mib r}'
\Psi_\sigma^\dagger({\mib r})
\Psi_{\sigma'}^\dagger({\mib r}')
U({\mib r}-{\mib r}')
\Psi_{\sigma'}({\mib r}')
\Psi_\sigma({\mib r})
\label{eq.3.1}
\end{eqnarray} 
and
\begin{equation}
{\hat {\mib J}}({\mib r})={e \over 2m}\sum_{\sigma}
[\Psi_\sigma^\dagger({\mib r}){\hat {\mib p}}\Psi_\sigma({\mib r})
-
({\hat {\mib p}}\Psi_\sigma^\dagger({\mib r}))\Psi_\sigma({\mib r})]
-{e^2 \over m}{\mib A}({\mib r})\sum_\sigma
\Psi_\sigma^\dagger({\mib r})\Psi_\sigma({\mib r}).
\label{eq.3.2}
\end{equation}
Here ${\mib A}({\mib r})$ represents the vector potential.  
(We treat the vector potential classically in this paper.) 
We also define the spin magnetization operator
\begin{equation}
{\hat {\mib M}}_{\rm S}({\mib r})\equiv
-g\mu_{\rm B}\sum_{\sigma\sigma'}\Psi_\sigma^\dagger({\mib r})
{1 \over 2}{\mib \sigma}_{\sigma\sigma'}
\Psi_{\sigma'}({\mib r}),
\label{eq.3.6}
\end{equation}
where $g$ and $\mu_{\rm B}$ represent the $g$-value and the Bohr magneton, 
respectively.
\par
The vector potential ${\mib A}({\mib r})$ originates from the local spin 
magnetization ${\mib M}_{\rm S}({\mib r})
=\langle{\hat {\mib M}}_{\rm S}({\mib r})\rangle$ 
and the electric current ${\mib J}({\mib r}$). 
The magnetic flux density ${\mib B}({\mib r})$ and the magnetic field 
${\mib H}({\mib r})$ satisfy
\begin{equation}
\bigtriangledown\times{\mib A}({\mib r})
=
{\mib B}({\mib r})
=\mu_0{\mib H}({\mib r})+{\mib M}_{\rm S}({\mib r})
\label{eq.3.3}
\end{equation}
and the Maxwell equation
\begin{equation}
\bigtriangledown\times{\mib H}({\mib r})={\mib J}({\mib r}).
\label{eq.3.4}
\end{equation}
In eq.~(\ref{eq.3.3}), $\mu_0$ denotes the magnetic permeability 
of the vacuum. 
Under the Coulomb gauge, $\bigtriangledown\cdot{\mib A}({\mib r})=0$, 
we find, from eqs.~(\ref{eq.3.3}) and (\ref{eq.3.4}),
\begin{equation}
\bigtriangleup{\mib A}({\mib r})=-\mu_0{\mib J}({\mib r})-
\bigtriangledown\times{\mib M}_{\rm S}({\mib r}).
\label{eq.3.5}
\end{equation}
Once the wavefunction ($T=0$) or the density matrix ($T\ne 0)$ is specified, 
${\mib A}({\mib r})$ is determined by eq.~(\ref{eq.3.5}) with appropriate 
boundary conditions. 
\par
The energy of the electron system is obtained as the expectation value 
of eq.~(\ref{eq.3.1}). 
To obtain the total energy, we have to take the magnetic field energy 
also into account, which is given by
\begin{equation}
E_{\rm M}={1 \over 2}\int{\rm d}{\mib r}{\mib H}({\mib r})
\cdot{\mib B}({\mib r})={1 \over 2}\int{\rm d}{\mib r}{\mib A}({\mib r})
\cdot{\mib J}({\mib r}).
\label{eq.3.7}
\end{equation}
In the last part of eq.~(\ref{eq.3.7}), we have put the surface integration 
of ${\mib H}({\mib r})\times{\mib A}({\mib r})$ equal to zero by taking 
the surface much larger than the sample. 
(When the sample is isolated, the current must circulate in it, which 
looks like a magnetic dipole moment at a place far away from the sample. 
Then, since the corresponding vector potential and magnetic field behave 
as $\sim r^{-2}$ and $r^{-3}$, respectively ($r$: distance from the sample), 
the surface integration of their product disappears for an infinitely 
large surface.\cite{rfa})
\par
Exactly speaking, there is the Zeeman energy by the spin magnetization
\begin{equation}
E_{\rm Z}=-\int{\rm d}{\mib r}{\mib M}_{\rm S}({\mib r})
\cdot{\mib H}({\mib r}).\label{eq.3.8}
\end{equation}
However, the following analysis does not treat this effect. 
We will give a brief discussion about this point later. 
(Note that the Zeeman energy by the orbital magnetization due to 
the spontaneous current is included in eq.~(\ref{eq.3.1}) 
through the vector potential.)

\subsection{Zero temperature}
\hspace*{\parindent}
Let us start with the case at $T=0$. 
As discussed in $\S2$, we introduce $|\psi\rangle$, which has 
the local current, ${\mib J}_\psi({\mib r})$, the total current, 
${\mib J}_\psi$, the induced magnetic field described by 
${\mib A}_\psi({\mib r})$ and the local magnetization, 
${\mib M}_{{\rm S}\psi}({\mib r})$. 
We also introduce $|\phi\rangle$ defined by eq.~(\ref{eq.2.5}) 
with the local current, ${\mib J}_\phi({\mib r})={\mib J}_\psi({\mib r})
+\delta{\mib J}({\mib r})$, the vector potential, 
${\mib A}_\phi({\mib r})={\mib A}_\psi({\mib r})+\delta{\mib A}({\mib r})$ 
and the local magnetization, ${\mib M}_{{\rm S}\phi}({\mib r})$. 
Since the exponential factor in eq.~(\ref{eq.2.5}) does not change 
spins of electrons, the magnetization should be the same between 
the two states, $|\psi\rangle$ and $|\phi\rangle$. 
Then, from eq.~(\ref{eq.3.5}) for the sets
$({\mib A}_\psi({\mib r}), {\mib J}_\psi({\mib r}), 
{\mib M}_{{\rm S}\psi}({\mib r}))$ and 
$({\mib A}_\phi({\mib r}), {\mib J}_\phi({\mib r}), 
{\mib M}_{{\rm S}\phi}({\mib r}))$, one finds
\begin{equation}
\bigtriangleup\delta{\mib A}({\mib r})=-\mu_0\delta{\mib J}({\mib r}),
\label{eq.3.9}
\end{equation}
where the explicit form of $\delta{\mib J}({\mib r})$ is derived 
from the expectation value of eq.~(\ref{eq.3.2}) to be
\begin{equation}
\delta{\mib J}({\mib r})=
{e \over m}
\langle\psi|\sum_\sigma\Psi_\sigma^\dagger({\mib r})\Psi_\sigma({\mib r})
|\psi\rangle
[\delta{\mib P}-e\delta{\mib A}({\mib r})].
\label{eq.3.10}
\end{equation}
The term with $O(\delta P^2)$ is dropped in eq.~(\ref{eq.3.10}). 
Substituting eq.~(\ref{eq.3.10}) into eq.~(\ref{eq.3.9}) and 
putting $\delta{\mib A}({\mib r})\equiv\delta{\mib P}a({\mib r})$, 
we find that the equation of $a({\mib r})$ is independent of 
$\delta{\mib P}$; $\delta{\mib A}({\mib r})$ and $\delta{\mib J}({\mib r})$ 
are therefore estimated as $O(\delta{\mib P})$. 
Keeping it in mind, we henceforth drop terms with $O(\delta P^2)$, 
by putting $\delta{\mib P}$ small enough so that such terms can be 
neglected compared with the ones with $O(\delta{\mib P})$. 
\par
Difference between $|\psi\rangle$ and $|\phi\rangle$ 
in the field energy up to $O(\delta{\mib P})$ is
\begin{eqnarray}
E_{{\rm M}\phi}
&\equiv&
{1 \over2 }\int{\rm d}{\mib r}{\mib A}_\phi({\mib r})
\cdot{\mib J}_\phi({\mib r})
\nonumber\\
&=&
{1 \over2 }\int{\rm d}{\mib r}{\mib A}_\psi({\mib r})
\cdot{\mib J}_\psi({\mib r})
+\int{\rm d}{\mib r}
\delta{\mib A}({\mib r})\cdot{\mib J}_\psi({\mib r})
\equiv E_{{\rm M}\psi}+\delta E_{\rm M},
\label{eq.3.11}
\end{eqnarray} 
where we have put the surface integrations of 
$\delta{\mib A}({\mib r})
\times(\bigtriangledown\times{\mib A}_\psi({\mib r}))$ and 
${\mib A}_\psi({\mib r})\times(\bigtriangledown\times
\delta{\mib A}({\mib r}))$ equal to zero as noted before. 
Substitution of ${\mib J}_\psi({\mib r})$ into $\delta E_{\rm M}$ gives
\begin{equation}
\delta E_{\rm M}=
\langle\psi|
\int{\rm d}{\mib r}\delta{\mib A}({\mib r})\cdot
{e \over m}\sum_\sigma
\Psi_\sigma^\dagger({\mib r})
[
{\hat {\mib p}}-e{\mib A}_\psi({\mib r})
]
\Psi_\sigma({\mib r})
|\psi\rangle.
\label{eq.3.12}
\end{equation}
where we have used the constraint 
$\bigtriangledown\cdot\delta{\mib A}({\mib r})=
\bigtriangledown\cdot[{\mib A}_\phi({\mib r})-{\mib A}_\psi({\mib r})]
=0$.
\par
Calculation of the expectation value of the Hamiltonian is performed 
in the same manner as in $\S2$. 
Noting that ${\mib A}_\psi({\mib r})$ (${\mib A}_\phi({\mib r})$) has 
to be used in eq.~(\ref{eq.3.1}) in evaluating
 $\langle\psi|H|\psi\rangle$ ($\langle\phi|H|\phi\rangle$), we obtain
\begin{eqnarray}
\langle\phi|H|\phi\rangle
&=&
\langle\psi|H|\psi\rangle
+\delta{\mib P}\cdot
\langle\psi|
\int{\rm d}{\mib r}{1 \over m}\sum_\sigma
\Psi_\sigma^\dagger({\mib r})
[
{\hat {\mib p}}-e{\mib A}_\psi({\mib r})
]
\Psi_\sigma({\mib r})
|\psi\rangle
-\delta E_{\rm M}
\nonumber\\
&=&
\langle\psi|H|\psi\rangle
+{1 \over e}\delta{\mib P}\cdot{\mib J}_\psi
-\delta E_{\rm M},
\label{eq.3.13}
\end{eqnarray}
where
\begin{equation}
{\mib J}_{\psi}
=
\langle\psi|
\sum_\sigma\int{\rm d}{\mib r}
\Psi_\sigma^\dagger({\mib r})
{e \over m}[{\hat {\mib p}}-e{\mib A}_{\psi}({\mib r})]
\Psi_\sigma({\mib r})
|\psi\rangle.
\label{eq.3.16b}
\end{equation} 
{}From eqs.~(\ref{eq.3.11}) and (\ref{eq.3.13}), 
the relation between the total 
energy of $|\psi\rangle$ $(E_\psi)$ and that of $|\phi\rangle$ $(E_\phi)$ is
\begin{equation}
E_\phi=E_{{\rm M}\phi}+\langle\phi|H|\phi\rangle
=E_\psi+{1 \over e}\delta{\mib P}\cdot{\mib J}_\psi.
\label{eq.3.14}
\end{equation}
When $\delta{\mib P}$ is put opposite to ${\mib J}_\psi$ in direction 
and small enough so that dropped terms with $O(\delta P^2)$ are less 
important in comparison with the second one in eq.~(\ref{eq.3.14}), 
$|\phi\rangle$ always has a lower energy than $|\psi\rangle$; 
this means that any state with a finite total electric current is always 
accompanied by a state with a lower energy, and it cannot be the ground state. 
\par
The present theory does not deny the possibility that the Zeeman energy 
given by eq.~(\ref{eq.3.8}) stabilizes the spontaneous current state 
with finite net current. 
When this kind of mechanism really works, however, the system has 
a finite local spin magnetization, which should be observed experimentally. 
 
\subsection{Finite Temperature}
\hspace*{\parindent}
Discussion at finite temperature in the presence of the induced magnetic 
field is almost the same as that in section 2.3. 
The difference is that $E_\phi$ and $E_\psi$ are now given by 
\begin{eqnarray}
\left\{
\begin{array}{l}
E_{\psi}=\sum_iw_i
[\langle\psi_i|H({\mib A}_{\psi_i})|\psi_i\rangle+E_{{\rm M}\psi_i}],\\
E_{\phi}=\sum_iw_i
[\langle\phi_i|H({\mib A}_{\phi_i})|\phi_i\rangle+E_{{\rm M}\phi_i}].
\end{array}
\right.
\label{eq.3.15}
\end{eqnarray}
where $H({\mib A}_{\psi_i(\phi_i)})$ means that 
${\mib A}_{\psi_i(\phi_i)}({\mib r})$ is used in eq.~(\ref{eq.3.1}). 
Each term in eq.~(\ref{eq.3.15}) can be evaluated in the same way as 
in section 3.2, and we again reach eq.~(\ref{eq.2.19}), where ${\mib J}_\psi$ 
in the present case is given by
\begin{equation}
{\mib J}_{\psi}
=\sum_iw_i
\langle\psi_i|
\sum_\sigma\int{\rm d}{\mib r}
\Psi_\sigma^\dagger({\mib r})
{e \over m}[{\hat {\mib p}}-e{\mib A}_{\psi_i}({\mib r})]
\Psi_\sigma({\mib r})
|\psi_i\rangle.
\label{eq.3.16}
\end{equation} 

We thus arrive at the conclusion that {\it the spontaneous current 
which is finite in total cannot be stable thermodynamically even if 
the magnetic field induced by the current is taken into account, 
except for the case that coupling between the spin magnetization 
and the magnetic field stabilizes this state.}

\section{Summary}
\hspace*{\parindent}
In this paper, we have examined second quantized version of the Bloch theorem, 
which can explicitly cover the superconductivity, and applied it 
to the recently proposed spontaneous surface current. 
We have also extended the theorem to systems with (1) the magnetic field 
by the spontaneous current and (2) magnetic impurities. 
\par
Now, we summarize the results in this paper.
\begin{itemize}
\item[(1)] Any state with finite spontaneous total current is not stable 
at all temperatures, even if the magnetic field induced by the current 
is taken into account. 
This statement is correct unless the Zeeman energy by the spin magnetization 
stabilizes this state.
\item[(2)] The recently proposed spontaneous surface current, which is also 
finite in total, is not stable at all temperatures. 
(Note that the Bloch theorem does not exclude the possibility of the surface 
current itself.) If a state with surface current is realized, 
the theorem states the existence of another counter current inside the system, 
which cancels the surface one as a whole.
\end{itemize}

We will explicitly show in a subsequent paper that the surface current state 
really has higher energy than the one with zero total current.\cite{end1} 

%%%%%%%%%%%%%%%%%%%%%%%%%%%%%%%%%%%%%%%%%%%%%%%%%%%%%%%%%%%%%%%%%%%%%%%%
\section*{Acknowledgements}
We wish to thank Professor S. Takada for pointing out the importance of 
the Bloch theorem, reading this manuscript and giving us various comments. 
Thanks are due to Mr. M. Matsumoto for kindly teaching us his theory 
at the seminar in University of Tsukuba. 
One of the authors (Y.O.) would like to thank Professors K. Kubo, 
T. Soda, Dr. D. Hirashima and Mr. T. Mutou for discussions and comments. 
He is also indebted to Mr. H. Ikeda for sending him ref.18. 

%%%%%%%%%%%%%%%%%%%%%%%%%%%%%%%%%%%%%%%%%%%%%%%%%%%%%%%%%%%%%%%%%%%%%%%%
%\newpage

%%%%%%%%%%%%%%%%%%%%%%%%%%%%%%%%%%%%%%%%%%%%%%%%%%%%%%%%%%%%%%%%%%%%%%%%%%
\newpage
\begin{center}
{\bf Figure Captions}
\end{center}
\begin{itemize}
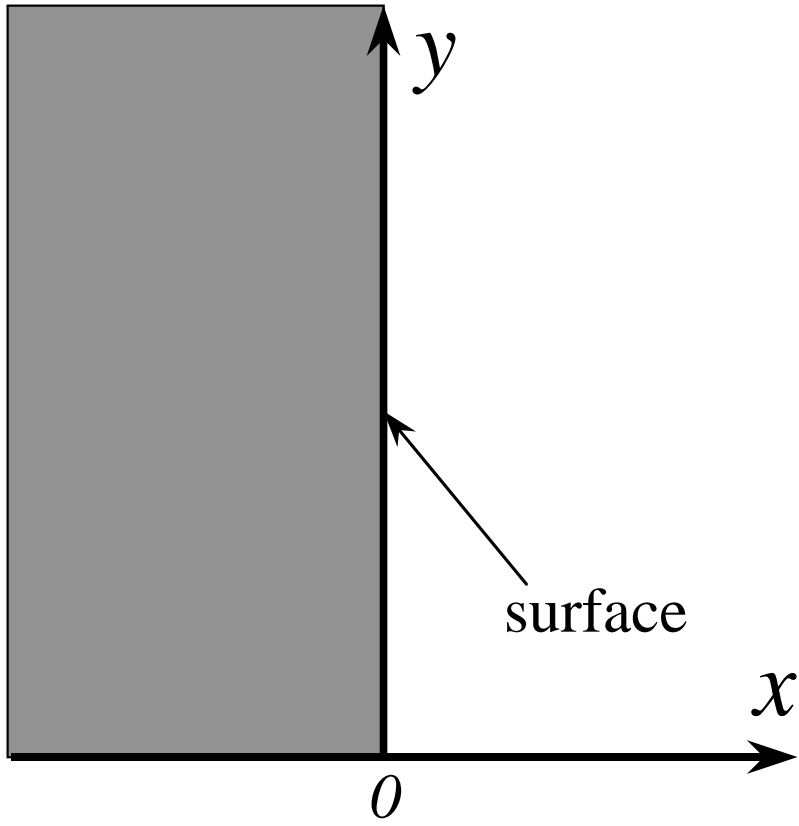
\item[Fig.~1:] Schematic picture of the semi-infinite system. 
$V({\mib r})=0$ in the right hand side in the figure and 
$V({\mib r})=\infty$ in the shaded region. 
In the figure, $z$-axis is perpendicular to the surface of the paper.
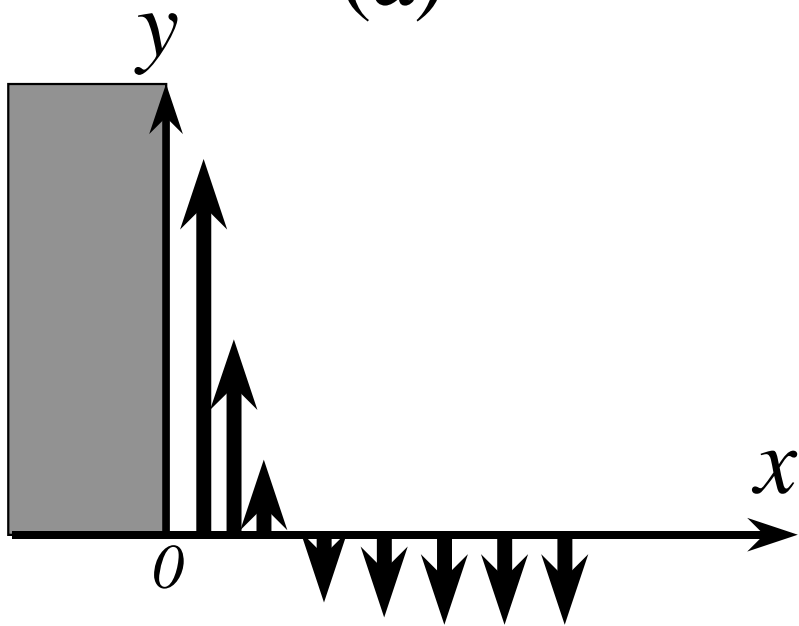
\item[Fig.~2:] (a) Assumed ground state, $|\psi\rangle$, which 
has a finite surface current. 
In the figure and also in the following figures, thick arrows 
represent the electric current. 
(b) A possible surface current state having a counter flow 
inside the system which cancels the former in total.
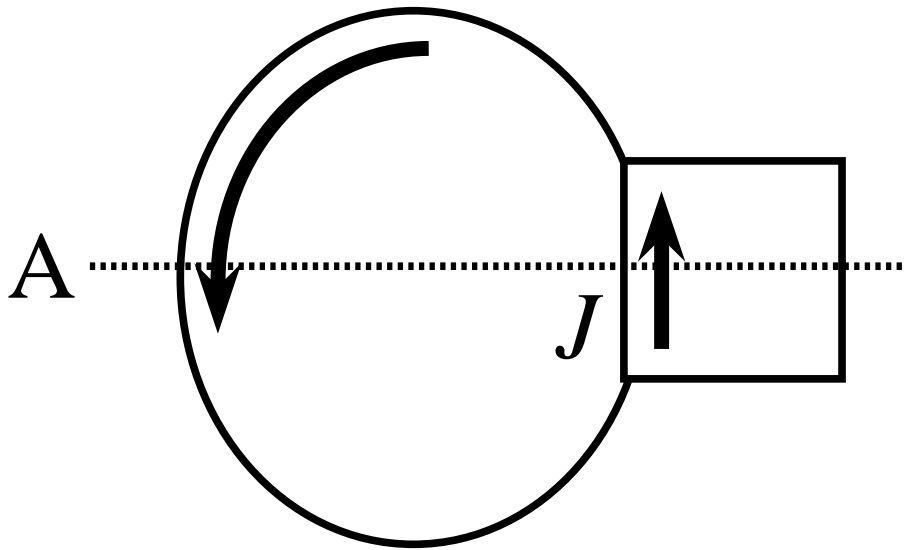
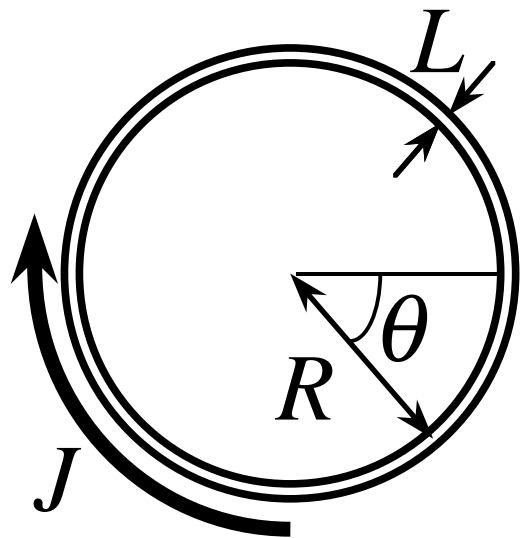
\item[Fig.~3:] (a) Circuit of spontaneous current. 
(b) A large ring with the width $L$ and the radius $R\gg L$. 
\end{itemize}

\begin{thebibliography}{99}
\bibitem{rf1} T.~Imai, T.~Shimizu, H.~Yasuoka, Y.~Ueda and K.~Kosuge: 
J.~Phys.~Soc.~Jpn.~{\bf 57} (1988) 2280.
\bibitem{rf2} P.~C.~Hammel, M.~Takigawa, R.~H.~Heffner, Z.~Fisk and 
K.~C.~Ott: Phys.~Rev.~Lett.~{\bf 63} (1989) 1992.
\bibitem{rf3} Y.~Kitaoka, S.~Ohsugi, K.~Ishida and K.~Asayama: 
Physica C{\bf 170} (1990) 189.
\bibitem{rf4} Y.~Ito, H.~Yasuoka, Y.~Fujiwara, Y.~Ueda, T.~Machi, 
I.~Tomeno, K.~Tai, N.~Koshizuka and S.~Tanaka: J.~Phys.~Soc.~Jpn.~
{\bf 61} (1992) 1287.
\bibitem{rf5} D.~A.~Wollman, D.~J.~Van Harlingen, W.~C.~Lee, D.~M.~Ginsberg 
and A.~J.~Legget: Phys.~Rev.~Lett.~{\bf 71} (1993) 2134; D.~A.~Wollman, 
D.~J.~Van Harlingen, J.~Giapintzakis and D.~M.~Ginsberg: {\it ibid}. 
{\bf 74} (1995) 797; D.~J.~Van Harlingen: Rev.~Mod.~Phys.~{\bf 67} (1995) 515.
\bibitem{rf6} C.~C.~Tsuei, J.~R.~Kirtley, C.~C.~Chi, L.~S.~Yu-Jahnes, 
A.~Gupta, T.~Shaw, J.~Z.~Sun and M.~B.~Ketchen: Phys.~Rev.~Lett.~{\bf 73} 
(1994) 593.
\bibitem{rf7} I.~Iguchi and Z.~Wen: Phys.~Rev.~B{\bf 49} (1994) 12388.
\bibitem{rf8} A.~Mathai, Y.~Gim, R.~C.~Black, A.~Amar and F.~C.~Wellstood: 
Phys.~Rev.~Lett.~{\bf 74} (1995) 4523.
\bibitem{rf9} K.~Kuboki and M.~Sigrist: J.~Phys.~Soc.~Jpn.~{\bf 65} (1996) 361.
\bibitem{rf10} M.~Sigrist, D.~B.~Bailey and R.~B.~Laughlin: Phys.~Rev.~Lett.~
{\bf 74} (1995) 3249.
\bibitem{rf11} M.~Matsumoto and H.~Shiba: J.~Phys.~Soc.~Jpn.~{\bf 64} (1995) 
3384.
\bibitem{rf12} M.~Matsumoto and H.~Shiba: J.~Phys.~Soc.~Jpn.~{\bf 64} (1995) 
4867.
\bibitem{rf13} M.~Matsumoto and H.~Shiba, preprint.
\bibitem{rf14} D.~Bohm: Phys.~Rev.~{\bf 75} (1949) 502.
\bibitem{rf15} H.~G.~Smith and J.~O.~Wilhelm, Rev.~Mod.~Phys.~{\bf 7} (1935) 
266.
\bibitem{rf16} S.~Takada and T.~Izuyama: Prog.~Theor.~Phys.~{\bf 41} (1969) 
635.
\bibitem{rf17} A.~Haug: {\it Theoretical Solid State Physics} 
(Pergamon Press 1972) Vol.1 Chap.II.A.
\bibitem{rf18} L.~Brillouin: Proc.~Roy.~Soc.~A{\bf 152} (1935) 19.
\bibitem{rfb} M.~Sigrist, T.~M.~Rice and K.~Ueda: Phys.~Rev.~Lett.~{\bf 63} 
(1989) 1727.
\bibitem{rfc} M.~Sigrist and K.~Ueda: Rev.~Mod.~Phys.~{\bf 63} (1991) 239.
\bibitem{end2} G.~Vignale: Phys.~Rev.~B{\bf 51} (1995) 2612.
\bibitem{rfa} H.~Takahashi: {\it Electro-Magnetism} (Shokabo 1959) Chap.4, 
[in Japanese].
\bibitem{end1} Y.~Ohashi and T.~Momoi: in preparation.
\end{thebibliography}
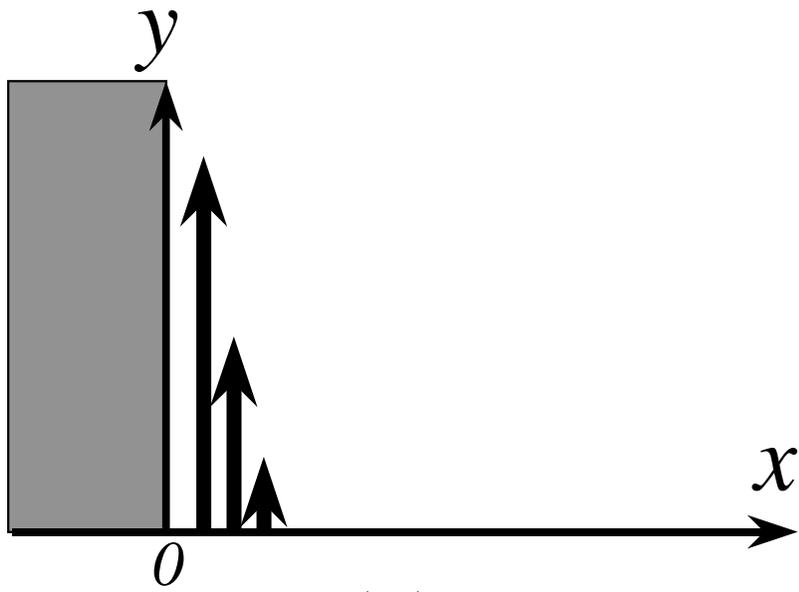
\end{document}